\documentclass[rmp,amsmath,amsbsy,twocolumn]{revtex4-2}
\usepackage{graphicx}
\usepackage{float}
\usepackage{epstopdf,cancel}
\usepackage{epsf,latexsym,bbm,euscript}
\usepackage{amssymb,amsmath,mathtools}
\usepackage{mathtools} 
\usepackage{times,graphics}
\usepackage{soul,xcolor}
\usepackage{mathtools}
\usepackage{mathrsfs}

\usepackage{enumitem}

\def\6{{\langle}}
\def\9{{\rangle}}

\newcommand{\lan}{\left\langle}
\newcommand{\ran}{\right\rangle}

\newcommand{\defeq}{\vcentcolon=}
\newcommand{\eqdef}{=\vcentcolon}

\newcommand{\be}{\begin{equation}}
\newcommand{\ee}{\end{equation}}
\newcommand{\ba}{\begin{eqnarray}}
\newcommand{\ea}{\end{eqnarray}}

\def\sg{\textsl{g}}

\def\rin{\mathrm{in}}

\def\rqm{\mathrm{qm}}
\def\rcl{\mathrm{cl}}
\def\rin{\mathrm{int}}

\def\half{{\tfrac{1}{2}}}
\def\tr{{\mathrm{tr}}}

\def\pad{{\partial}}

\def\hcl{{ \hat{\cL}}}

\def\sg{\textsl{g}}

\def\eK{\EuScript{K}}
\def\eH{\EuScript{H}}
\def\eL{\EuScript{L}}

\def\eP{\EuScript{P}}

\def\cA{\mathcal{A}}
\def\cB{\mathcal{B}}
\def\cF{\mathcal{F}}

\def\cL{\mathcal{L}}

\def\cS{\mathcal{S}}

\def\mbr{\mathbb{R}}
\def\mbc{\mathbb{C}}

\def\gA{\mathfrak{A}}

\def\gacl{\mathfrak{A}_\mathrm{cl}}
\def\gaqm{\mathfrak{A}_\mathrm{qm}}

\def\rhcl{\rho_\mathrm{cl}}
\def\rhqm{\rho_\mathrm{qm}}

\def\hcl{H_\mathrm{cl}}
\def\hqm{H_\mathrm{qm}}
\def\hin{H_\mathrm{int}}
\def\tP{\mathtt{P}}

\def\hbpi{\hat{\boldsymbol{\Pi}}}
\def\bca{\boldsymbol{\mathcal{A}}}
\def\vxi{{\boldsymbol\xi}}

\def\vH  {{\boldsymbol{\xi}_H}}
\def\veta{{\boldsymbol{\eta}}}

\def\hA{{\hat{A}}}
\def\hH{{\hat{H}}}
\def\hek{\hat{\eK}}
\def\hel{\hat{\eL}}

\def\hbz{\hat{\boldsymbol{z}}}
\def\bj{ {\boldsymbol{j}}}

\def\aH{\mathsf{H}}

\usepackage{url,hyperref}
\hypersetup{colorlinks,linkcolor={blue!55!black},citecolor={red!50!black},urlcolor={blue!45!black}}

\makeatletter
\newcommand{\ostar}{\mathbin{\mathpalette\make@circled \star}}
\newcommand{\make@circled}[2]{%
  \ooalign{$\m@th#1\smallbigcirc{#1}$\cr\hidewidth$\m@th#1#2$\hidewidth\cr}%
}
\newcommand{\smallbigcirc}[1]{%
  \vcenter{\hbox{\scalebox{0.77778}{$\m@th#1\bigcirc$}}}%
}
\begin{document}

	\title{Classical-Quantum Hybrid Models}
\author{Daniel R. Terno}
\affiliation{School of Mathematical and Physical Sciences, Macquarie University, Sydney, New South Wales 2109, Australia}
\email{daniel.terno@mq.edu.au}	

\begin{abstract}
Hybrid classical-quantum models are computational schemes that investigate the time evolution of systems, where some degrees of freedom are treated classically, while others are described quantum-mechanically. First, we present the motivation for such models, outline the requirements they must satisfy, and provide explanations for their development. Then we review various popular non-relativistic schemes and their associated limitations, with a particular emphasis on reversible dynamics.



\end{abstract}	
	\maketitle
\tableofcontents 	
	\section{Introduction}

 Phenomena described by hybrid classical-quantum models range from   interactions of elementary particles with a superheated liquid in a bubble chamber to the inflaton field driving the expansion of the Universe (Boucher and Traschen, 1988). There are several reasons for adopting a hybrid formalism.

At the purely pragmatic level, it provides an efficient computational scheme. For instance, investigations of atomic and molecular dynamics require simulating the time evolution of increasingly large electronic and nuclear complexes. The exponentially high computational costs of such simulations can be managed through various strategies. One way to moderate the growth of required computational resources is to freeze the nuclear dynamics, statistically describe inner-shell populations, and explicitly treat only several chemically active electrons per atom (Fulde, 2012). Alternatively, the full set of degrees of freedom can be retained by selecting a subsystem that is described quantum-mechanically while the rest of the degrees of freedom is treated classically (Tully, 1998; Crespo-Otero and Barbati, 2018). The selection criteria typically involve the separation into fast and slow degrees of freedom in the Born--Oppenheimer approximation (Bayer, 2006). Similarly, numerous processes in condensed matter systems find convenient descriptions through classical or thermodynamic language, with quantum mechanics providing values for a limited set of parameters. While the influence of this classically described background on quantum systems is, in principle, well-understood, dealing with back reaction --- where the quantum system affects its environment --- presents a considerably more complex problem. Hybrid mechanics is specifically designed to address this challenge.

At the fundamental level, the emergence of the classical world as a limit of a quantum description is still not fully understood (Zurek, 2003; Schlosshauer, 2007). One aspect of this problem involves understanding the classical limit of only one part of the system, which necessitates a consistent dynamics for the resulting quantum-classical hybrid. Moreover, regardless of one's preferred interpretation of quantum theory, the outcome of measurement is a classical record (Peres, 1995; Busch et al., 2016). Therefore, achieving a consistent description of the quantum and classical sectors engaged in mutual interaction is essential for the logical coherence of presenting quantum foundations.

Finally, despite decades of research, a fully developed quantum theory of gravity remains elusive, and there is a notable absence of compelling experimental evidence for its quantum signatures (Kiefer, 2012). Therefore, we should seriously consider an old hypothesis suggesting that the gravitational field is classical, even though its material sources are quantized. The success or failure of a hybrid theory that describes classical gravity and quantum matter, as well as the form it eventually takes, will significantly influence our understanding of their interaction (Oppenheim, 2023,2023b).

Different purposes of various hybrid models lead to different expectations. Effective computational schemes just need to be ``good enough" for the duration of the investigated process, even if it is desirable to be assured of their consistency. Candidates for a fundamental theory ought  to satisfy a number of consistency requirements. Successes and failures in reaching this goal form the subject matter of this chapter.
\subsection{Scope and structure}

First we summarise the necessary formal aspects of classical (Sec.~\ref{cm}) and quantum (Sec.~\ref{qm}) theories. The   constraints on hybrid models that follow from their compatibility with the rest of the known physics as well as from the perspective of generalised probability theories (GPTs) are discussed in Sec.~\ref{axioms}.

  By borrowing terminology from quantum theory, hybrid models can be broadly separated into  reversible (unitary) and irreversible (completely positive) schemes. Since most of the literature is devoted to the reversible schemes, they are our primary focus. After reviewing the mean field models (Sec.~\ref{mean}), we describe the algebraic aspects of constructing   hybrid brackets in Sec.~\ref{hb}. Intuitively, hybrid mechanics should involve objects of the same nature.  Hence many attempts to write hybrid schemes start from rewriting   classical mechanics  in terms of wave functions, or reformulating   quantum mechanics on the phase space.  Hilbert space constructions avoid various no-go theorems of Sec.~\ref{hb} and are described in Sec.~\ref{hspace}. Phase space methods and their role in deriving the classical limit and hybrid schemes are discussed in Sec.~\ref{ps}. After reviewing the schemes of reversible dynamics and their problems, the irreversible hybrid dynamics is discussed in Sec.~\ref{stoc}. We  introduce schemes that incorporate decoherence and diffusion to ensure  compliance with the minimal consistency requirements of the hybrid models.

The discussion is necessarily concise. We work in the non-relativistic regime and, aside from a brief description of the semiclassical Einstein equations, do not discuss quantum or classical fields, as well as fermions and the anti-commutation relations. No position on any controversy in the foundations of quantum mechanics is taken. Only a short list of references is provided, but several monographs and topical reviews that together contain a nearly exhaustive survey of the literature are highlighted.

\subsection{Notation and conventions}

For brevity we refer to a quantum  system as Q and to a classical system as C. Generalized position and momentum are combined into  $z=(x,k)$, and the collection of derivatives is denoted as $\nabla_z=(\pad_x,\pad_k)$.
\begin{table}[!htb] \centering
	\begin{tabular}{|c |l|} \hline \small  		
    \begin{tabular}{@{}c@{}} \hspace{1mm} symbol \\
		\end{tabular}
				&
		\begin{tabular}{@{}c@{}} {definition} \\
		\end{tabular}
	\\ \hline \hline
    $ \gacl, \gaqm, \gA $ & algebras of classical, quantum, and hybrid operators \\ \hline
    $\eP$ & phase space (Poisson manifold) \\ \hline
    $ J$ & symplectic matrix \\ \hline
     $ I$ & identity operator or matrix \\ \hline
    $\vH $ & Hamiltonian vector field \\ \hline
     $ z=(x,k)$ & a set of $2n$ canonical positions and momenta \\ \hline
    $\mu(dz)$ & phase space measure   \\ \hline
	$\{f,g\}\equiv f \mathtt{P} g $ & Poisson bracket of  functions $f(z)$ and $g(z)$ \\ \hline
    $[A,B]$  &  commutator of the operators $A,B\in\gaqm$ \\ \hline
	$\{A,B\}_\hbar$ &  Poisson bracket of the operators, $[A,B]/i\hbar$ \\ \hline
    $[\![f,g]\!]$ & Moyal bracket of functions $f(z)$ and $g(z)$ \\ \hline
    $\{\![Y,F]\! \}$ & a  hybrid bracket of $Y,F\in\gA$ \\ \hline
    $W$ & Wigner function \\ \hline
    $\Upsilon(q,x,p,t)$ & quantum-classical wave function \\ \hline
     $\Upsilon(q,t)$ & quantum-classical wave function, Sec. VC\\ \hline
     $\varrho$ & hybrid density operator \\ \hline

	\end{tabular}
	\caption{\small Table of symbols recurrently used in the text.} 	\label{Tab:Signs} \vspace{-2mm}
\end{table}

We use carets to indicate operators only when they act on a concrete Hilbert space.  We denote a joint operation of the operator trace and integration over the phase space variables as
\be
\mathrm{Tr} A=\int\!\tr A\mu(dz).
\ee
The hybrid density operator $\varrho$ satisfies $\mathrm{Tr}\varrho=1$. The Einstein summation convention over repeated indices is assumed.

\section{Formal structures in mechanics} \label{formal}

Several structural components of classical and quantum theories (Arnold, 1989; Landsman, 1998; 2017)   are particularly important in construction of hybrids. In their description we follow the same pattern, emphasising the algebraic structures and the state--observable duality.   We begin with classical mechanics and introduce the shared concepts in its setting.

\subsection{Classical mechanics}\label{cm}

States and observables of a classical system are described  with the help of the phase space $\eP$. We restrict the discussion to non-constrained systems with finite number of degrees of freedom. Then the phase space is a  symplectic manifold that is a cotangent bundle of the configuration space. Local coordinates $z=(x,k)$ are formed from the generalized coordinates of the configuration space $x=\{x^a\}$ and the generalized momenta. These are related to the coordinates $x$ and velocities $\dot x$ via $k_a\defeq\pad L/\pad \dot x^a$, where $L(x,\dot x)$ is the system's Lagrangian.

Classical observables  are smooth functions on the phase space and form the  algebra $\gacl$. The Poisson bracket
\begin{align}
\{f,g\}&=\left(\frac{\pad f}{\pad x^a}\frac{\pad g}{\pad k_a}-\frac{\pad f}{\pad k_a}\frac{\pad g}{\pad x^a}\right)\nonumber \\
&\equiv
f\big(\overset{\leftarrow}{\partial_x}\overset{\rightarrow}{\partial_k}-\overset{\leftarrow}{\partial_k}\overset{\rightarrow}{\partial_x}\big)g\equiv f \mathtt{P} g,
\end{align}
 where arrows indicate direction of action of the differential operators,
governs the dynamics of observables  via the canonical equations of motion
\be
\dot x=\{x,H\}, \qquad \dot k=\{k,H\},
\ee
that are generated by the system's Hamiltonian. An arbitrary phase space function $f$ evolves according to
\be
\frac{df }{dt}=\frac{\pad f}{\pad t} +\{f,H\}. \label{classE}
\ee

A non-degenerate closed 2-form that determines the symplectic structure on $\eP$ can always be written in local coordinates as
\be
\omega^{(2)}=dp_a\wedge dq^a\equiv dp\wedge dq.
\ee
The symplectic form establishes  the isomorphism between vectors and 1-forms on $\eP$ by matching a vector $\veta$ with the form $\omega^{(1)}_\veta$ via $\omega^{(1)}_\veta(\boldsymbol{\xi})\defeq\omega^{(2)}(\veta,\boldsymbol{\xi})$. Hence in the local coordinates the Hamiltonian vector field is given by
\be
\vH=J\nabla_z H,
\qquad J= \begin{pmatrix}0 & I\\
-I & 0\end{pmatrix},
\ee
where $J$ is the symplectic matrix. The canonical equations thus become
\be
\dot z(t)=\vH\big(z(t)\big),
\ee
representing the Hamiltonian phase flow. Then
\be
 \{f,g\}=-\omega^{(2)}\big(\nabla_z f,\nabla_z g\big)=(\nabla_z f)^T\cdot J\cdot \nabla_z g.
\ee

 The Poisson bracket  is defined more abstractly  as a Lie bracket on the underlying manifold: it is linear, antisymmetric
and satisfies the Jacobi identity
\begin{align}
 \{f,\{g,h\}\}=\{\{f,g\},h\}+\{g,\{f,h\}\}.  \label{l2}
\end{align}
In addition it satisfies the Leibnitz rule with respect to the
product defining the algebra, $f\circ g (x,k)\defeq f(x,k)g(x,k)$,
\be
\{f,gh\}=\{f,g\}h+g\{f,h\}. \label{ll}
\ee
Technically this is the Jordan--Lee algebra with   associative multiplication, i.e. the Poisson algebra.

The  state space $\cS$ for an algebra $\gA$  with identity $I$ consists of all linear functionals  $\omega:\gA\to\mbc$ that are positive, ($\omega(A^*A)\geqslant0$ $\forall A\in\gA$) and normalized, $\omega(I)=1$. The states are continuous functionals.  $\cS(\gA)$ is a convex set, i. e. for two states $\omega_1,\omega_2$ and $0\leqslant\lambda \leqslant 1$ the mixture $\omega=\lambda\omega_1+(1-\lambda)\omega_2$ belongs to $\cS$, which is a closed subset of the unit ball in the dual space of the algebra, $\gA^*$. In classical mechanics the states are described by the Liouville probability density  $\rhcl$, with
\begin{align}
&\rhcl\geqslant0, \qquad \int \rhcl(x,k,t) \mu(dxdk)=1, \label{rcl1} \\
& \6A\9_\rho= \int A(x,k,t) \rhcl(x,k,t) \mu(dxdk), \label{rcl2}
\end{align}
where the measure $\mu(dxdk)$  follows from the volume form $\omega^{(2)^n}$ on the phase space. For $\eP=\mbr^{2n}$ the measure is  $\mu(dz)=d^nxd^nk$   abbreviated as $dxdk$. Pure states of a classical system are the points of its phase space $\eP$, and the corresponding Liouville density is a distribution
\be
\rhcl^{z_0}=\delta(x-x_0)\delta(k-k_0).
\ee

The convex set of all probability measures on a  topological space
is a generalized simplex. Its vertices are all point-concentrated
measures. It is an important property of the simplexes that each point
of a simplex can be uniquely represented as a convex combination (finite or infinite) of the extremal
points. This uniqueness of decomposition is a crucial distinction between classical and quantum mixtures (Mielnik, 1974).

Hamiltonian flows preserve the symplectic structure and the variety of the derived invariants, including the phase space volume. Invariance of the overlaps of the Liuoville densities --- the classical unitarity ---
 is expressed as the Liouville equation,
\be
\pad_t \rhcl=-\{\rhcl,H\}. \label{Liouville}
\ee

\subsection{Quantum mechanics}\label{qm}

Quantum observables are elements of the self-adjoint part $\gaqm$ of the relevant operator algebra.
 A commutative product $\circ$ is introduced via
\be
A\circ B\defeq\half (AB+BA).
\ee
 Quantum Poisson bracket is defined as
\be
\{A,B\}_\hbar=\frac{1}{i\hbar}(AB-BA).
\ee
It is a Lie bracket and a derivation in $\gaqm$. In analyzing the classical limit the Planck constant is treated as a variable parameter and the limit $\hbar\to 0$ is studied.

The associator identity
\be
(A\circ B)\circ C-A\circ (B\circ C)=\tfrac{1}{4}\hbar^2\{\{A,C\},B\},
\ee
is what differentiates the commutative product $\circ$ from its classical counterpart. When $\circ$ is associative the algebra becomes the Poisson algebra, and it formally corresponds to $\hbar\equiv 0$.

In the Heisenberg picture operators evolve
\be
\frac{d A^\aH}{dt}=\{A^\aH,H\}_\hbar+\frac{\pad A^\aH}{\pad t}.
\ee

 In the Hilbert space representation the elements of $\cS$   are trace class trace one positive operators (density operators), and the expectation value of $A\in \gaqm$  is given by
\be
\6A\9_\rho=\tr(\rho A). \label{rhotr}
\ee
Pure states are elements  of the projective space $\psi\in \mathbbm{P}\eH$.
In the Schr\"{o}dinger picture   pure states evolve according to
\be
i\hbar\frac{d\psi}{dt}=H\psi,
\ee
and the density operator $\rho$ evolves as its classical counterpart,
\be
\pad_t\rho=-\{\rho,H\}_\hbar. \label{rhqme}
\ee

States and operators can be represented not only on  the Hilbert space  of square-integrable functions on the configuration space of the system, such as $\eH=\mathtt{L}^2(\mbr^n)$.
In many applications, such as quantum optics or quantum information with continuous variables (Weedbrook et al., 2012), representation of the state of a quantum system on the phase space of its classical counterpart is particularly convenient. The Wigner quasi-probability distribution function  (Hillary et al., 1984; Zachos et al., 2005),
\be
W_\rho(q,p)=\frac{1}{(2\pi\hbar)^n}\int\! dy e^{i p y/\hbar}\6q+\half y|\hat{\rho}|q-\half y\9,
\ee
represents the state $\rho$. The quasi-probability function  is normalized according to
\be
\int\!W_\rho(z)\mu(dz)=1, \qquad \tr \big(\hat A\hat\rho\big)=\int\!W_\rho(z)A(z)\mu(dz),
\ee
where $z=(q,p)$, but is not necessarily  positive-definite.

 A Gaussian state ${\rho}$ has a Gaussian characteristic function. Its Fourier transform  gives us a Gaussian Wigner function
\be
W_\rho(z)=\frac{\exp \left[-\half(z-\6 z\9)^T{\sigma}^{-1}(z-\6 z\9)\right]}{(2\pi)^n \sqrt{ \mathrm{det}\sigma}   },
\ee
 where $\sigma$ is a covariance matrix, namely,  matrix of  the second moments of the state $\hat{\rho}$.
By definition, a Gaussian  probability distribution  can be completely described by its first and second moments;  all  higher moments can be derived from the first two using the following method
\begin{align}
&\lan (z-\6 z\9)^k\ran=0~~~~~~~~~~~~~~~~~~~~~~~~~~~~~~\text{for odd}~ k,\\
&\lan (z-\6 z\9)^k\ran=\sum{(c_{ij}...c_{xz})}~~~~~~~~~\text{for even}~ k
\end{align}
 also known as Wick's theorem.
The sum is taken over all the different permutations of $k$ indices. Therefore we will have $(k-1)!/(2^{k/2-1}(k/2-1)!)$ terms where each consists of the product of $k/2$ covariances $c_{ij} \equiv \lan (z_i-\6 z\9_i)(z_j-\6 z\9_j)\ran$. Gaussian operations preserve the Gaussian character of the states they are applied to. However, not all states with $W_\rho>0$ can be prepared  using only Gaussian
states and operations (Filip and Mi\v{s}ta Jr., 2011).

Dynamics of an open system is often conveniently expressed via Gorini, Kossakowki, Sudarshan, Lindblad (GKSL) equation (Ingarden et al., 1997; Schlosshauer, 2007), one of whose forms is
\be
i\hbar\frac{d\rho}{d t}=-[\rho,H]+\half d^{\alpha\beta}\left([L_\alpha,\rho L_\beta^\dag]+[L_\alpha\rho, L_\beta^\dag]\right).
\ee
The Lindblad operators $L_\alpha$ are obtained from the interaction term, and the coefficients $d^{\alpha\beta}$ encapsulate all information about the
physical parameters of the decoherence and dissipation processes.

\subsection{Hybrid mechanics}\label{axioms}
A   consistent hybrid dynamics has to satisfy a number of restrictions that follow from it being a part of a broader physical picture. The following is the list  of requirements, of various degrees of importance and acceptance, that may or should be imposed on proposed hybrid schemes. It largely follows the list of Boucher and Traschen (1988). 

The most basic purpose of any hybrid scheme is to obtain predictions about the Q and C subsystems. It has to
 to identify the classical  and the quantum  sectors, as well as to produce
\begin{enumerate}[label=\Roman*.]\vspace{-1mm}
\item a phase-space probability density $\rhcl$ that satisfies Eqs.~\eqref{rcl1} and \eqref{rcl2};
\vspace{-2mm}
\item a positive-semidefinite density operator $\rhqm$ that satisfies Eq.~\eqref{rhotr}. 
 \end{enumerate}

\noindent Each of the sectors behaves in the usual way, i.e.
\begin{enumerate}[label=\Roman*.]\setcounter{enumi}{2}  \vspace{-1mm}
\item if C and Q are uncoupled, then $\rhcl$ evolves according to Eq.~\eqref{Liouville} and $\rhqm$ according to Eq.~\eqref{rhqme}; \vspace{-2mm}
\item  classical canonical transformations and quantum unitary transformations are realised on C and Q sectors, respectively (equivariance).  \vspace{-1mm}
\end{enumerate}

While the evolution of the QC system may be unitary or not, and  have only one of the Schr\"{o}dinger or the Heisenberg pictures accessible,  it should to
\begin{enumerate}[label=\Roman*.]\setcounter{enumi}{4}  \vspace{-1mm}
\item conform to the laws of physics and at the very least (a) satisfy the standard conservation laws, particularly energy; (b) maintain impossibility of  superluminal communications; (c) conform to the second law of thermodynamics.
\end{enumerate}
\noindent These requirements have far-reaching consequences. In quantum mechanics,  non-linear evolution  is compatible with the principle  of superposition  and simply implies that a time-evolved initial pure state $\psi_1(0)+\psi_2(0)$ is different form $\psi_1(t)+\psi_2(t)$) (Peres, 1995). However, it enables one to distinguish non-orthogonal states. In turn, this leads to superluminal communication (Gisin, 1990) and to violation of the second law of thermodynamics (Peres, 1989).

One of the motivations for hybrid dynamics is to introduce back reaction of Q on C. The CQ correlations may be generated if
\begin{enumerate}[label=\Roman*.]\setcounter{enumi}{5}  \vspace{-1mm}
\item the quantum purity $\tr (\rhqm^2)$ is not a constant of motion.  \vspace{-2mm}
\end{enumerate}
This decoherence property (Gay-Balmaz and Tronci, 2022)  for initially pure $\rhqm$ under overall unitary evolution indicates building up of entanglement between C and Q systems.

As the minimal goal is to produce reasonable probability distributions, then having the Heisenberg picture, i. e., explicit evolution equations for all the observable operators can be described as a desirable feature. However, in this case it is reasonable to demand that at least
 \begin{enumerate}[label=\Roman*.]\setcounter{enumi}{6}  \vspace{-1mm}
 \item if purely classical and purely quantum (Heisenberg) equations of motion have the same form, the hybrid equations have the same form as well (Peres and Terno, 2001). More generally, the difference between the quantum and hybrid equations of motion and/or   classical unobservable quantities should disappear in the formal limit $\hbar\to 0$.
 \end{enumerate}
 Failure to comply with this form of the correspondence principle leads to a breaking down of the classical limit (taken for the system Q) when expressed via the statistical moments of the classical and quantum probability distributions (Terno (2006)).  

For illustration of various hybrid schemes we consider Hamiltonian systems (that   have a classical   Hamiltonian $H(x,k,q,p)$, or a quantum Hamiltonian $H(X,K,Q,P)$. Separating the hybrid Hamiltonian into C, Q and the interacting parts, we write
$H=\hcl(x,k)+\hqm(q,p)+\hin$. Hybrid schemes give the concrete meaning to this formal expression.

Two model will be routinely used to illustrate the hybrid schemes. One  example is provided by
two bilinear coupled oscillators, where masses and frequencies are absorbed into the definitions of canonical variables, thus making $\hbar$ dimensionless. Its classical version is
\be
H=\half x^2+\half k^2+\half q^2+ \half p^2+ \lambda xq, \label{ex1}
\ee
where in the hybrid picture $x,k$ remain classical operators, $\{x,k\}=1$ while the other canonical pair is promoted to operators $\{q,p\}_\hbar=1$.
  Another popular system consists of a classical particle and a quantum spin, with the interaction term
  \be
  \hin=\lambda\sigma_zk,
  \ee
where $\sigma_z$ is the Pauli $z$-matrix.

Several general results indicate that construction of the self-consistent hybrid dynamics modifies at least some of our expectations. Adapting the argument of Di\'{o}si et al. (2000), we conclude that no matter what form hybrid dynamics takes, it should be impossible from the measurement of $x$ and $k$ to obtain precise information about the quantum state. Hence, even if the Liouville distribution were introduced into phase space as a concession to statistical mechanics, some sort of epistemic  restrictions on classical information is inevitable (Bartlett et al., 2012).

Recent work of Galley et al. (2023) provides some indication of the possibilities for hybrid dynamics under the most relaxed requirements. Generalized probabilistic theories (GPTs),   provide a unified meta-theoretical framework in which rules of classical and quantum theory are  special cases (Janotta and Hinrichsen, 2014). Their primary focus is on the probabilistic relationships between preparation and effects and identification of the state space structure from the experimental results. For finite dimensional systems the exact ingredients of what are the necessary assumptions  to recover quantum or classical theory are quite well understood (Brandford et al., 2018). In particular, the discrete structure of classical pure states (Sec.~\ref{cm}) is reflected in identification of classical state spaces as simplexes.

Galley et al. (2023) considered   interacting classical (C) and non-classical (S) systems. If one requires that interaction leads to the precisely defined information flow from S to C (or its backreaction on C), and reversibility of that interaction, then a contradiction is established.

We now survey several of the most popular classes of models. A fully consistent mixed dynamics (Di\'{o}si et al. 2000) involves treating the entire Hamiltonian $H(X,K,Q,P)$ as quantum, following its evolution and then taking a partial classical limit. Such an approach is impractical, but truncation of the exact equations at a particular order leads to many of the schemes that are surveyed below.


\section{Mean-field models}\label{mean}
Known as mean--field or Ehrenfest models (Boucher and Traschen, 1988; Tully, 1998; Kapral, 2006; Crespo-Otero and Barbati, 2018) this scheme in its   basic  form evolves the classical variables $z=(x,k)$ and the wave function $\psi(t,q;x,k)\equiv\6q|\psi(t;z)\9$ as
\begin{align}
&\dot x=\{x,\6\hat{H}\9\}=\pad_k \6\hat{H}\9, \qquad \dot k=\{k,\6\hat{H}\9\}=-\pad_x\6\hat{H}\9, \label{zM} \\
&i\hbar\frac{d\psi}{dt}=i\hbar\left(\pad_t\psi+\6\vxi_\hH\9\cdot\nabla_z\psi\right)=\hat{H}\psi. \label{psM}
\end{align}
Here the classical evolution is driven by the expectation value of the hybrid Hamiltonian, $\6\hH\9=\6\psi_\rqm|\hH|\psi_\rqm\9$, and the phase space Hamiltonian vector $\6\vxi_\hH\9=(\pad_k\6\hH\9,-\pad_x\6\hH\9)$
enters the Schr\"{o}dinger equation through the dependence of the wave function on classical variables.

 The scheme satisfies requirements  I-IV, as well as the conservation of energy (Tully 1998; Manfredi et al., 2023). Its observables are Hermitian operator valued functions on phase space, and the scheme often provides accurate quantum transition probabilities, and can be augmented by additional terms or computational methods (Crespo-Otero and Barbati, 2018). Spin degrees of freedom can be naturally incorporated (Husain et al., 2022).
 The standard semiclassical Einstein equations
 \begin{align}
& G_{\mu\nu}(\sg)=\frac{8\pi G}{c^4}\6\psi|\hat{T}_{\mu\nu}(\hat\phi,\hat\pi;\sg)|\psi\9_\mathrm{ren}, \\
& i\hbar \dot\psi=\hH [\hat\phi,\hat\pi;\sg]\psi,
 \end{align}
 where the Einstein tensor is sourced by the renormalized energy-momentum tensor of the matter (fields $\hat\phi$ and their canonical momenta $\hat\pi$), and the quantum state of the matter fields $\psi$ is driven by the
 Hamiltonian that depends on the classical metric $\sg_{\mu\nu}$, is  one the more famous examples (Kiefer, 2012).

 Its drawbacks follow from its advantages. Absence of correlations between quantum and classical degrees of freedom is built into the scheme. Introducing the density operator $\hat{\rho}(t;z)$, $\mathrm{Tr}\hat\rho=1$ (with a view of replacing the sharp classical data with $\rho(t,x,k)=\tr\hat{\rho}$)  results in the Liouville-like equation
\be
\pad_t\hat{\rho}=-\{\hat{\rho},\hH\}_\hbar-\{\hat{\rho},\hH\}.
\ee
Its modifications, at least partially, allow  to introduce quantum--classical correlations (Alonso et al., 2012).

The scheme also does not satisfy the principle of detailed balance, which
means that at equilibrium a forward process is not balanced by its
reverse process. The inclusion of quantum corrections may also produce
Boltzmann distributions in the long-time limit (Crespo-Otero and Barbati, 2018).

From a fundamental perspective the main problem of this scheme is its nonlinearity. Because of the average description of the potential, th eformalism cannot successfully handle Q being in  superposition, particularly if the states are macroscopically distinct (Penrose, 1996). In general, evolution of a observable $\hA$ in the Ehrenfest model follows (Salcedo, 2012)
\be
\frac{d\6\hA\9_{\hat\rho}}{dt}=\int\!\left( \tr\big(\hat{\rho}\{\hA,\hH\}_\hbar\big)+\{\tr(\hat{\rho}\hA),\tr(\hat{\rho}\hH)\}\right)\mu(dz),
\ee
which clearly prevents   evolution of the expectation value of $A$ on some initial mixture into a mixture of evolved expectation values of the individual components of the mixture. As a result, the scheme allows one to discriminate between different  mixtures that realise the state $\hat{\rho}$. This  not only  violates  the Requirements Vb and Vc, but also demonstrates the internal inconsistency of the scheme.
Indeed,  if this scheme is used to describe the measurement process
in quantum mechanics, then the classical measurement apparatus would be able to perform operations that are forbidden in quantum theory, whose basic description is predicated on using the classical apparatus.

From a computational perspective the scheme can be successfully used if  (a) the bipartite CQ system starts in the product state; (b) he coupling between the subsystems is weak; (c) C is effectively semiclassical (i. e. the characteristic action $S_\rcl\gg\hbar$). Under these conditions the mean-field dynamics  is qualitatively similar to the full quantum evolution holds for a finite period of time and in some cases closely approximates the exact dynamics (Husain et al., 2023).

\section{Hybrid brackets} \label{hb}

Given that both classical and quantum dynamics are expressed with their respective Poisson brackets, one approach to the construction of hybrid schemes is the  construction of a bracket $\{\![\cdot,\cdot]\!\}$ that defines
the hybrid operator algebra on the tensor product space  $\gA=\gacl\otimes\gaqm$. Its elements   are linear combinations of products of  functions on the phase space
with quantum operators. Hence a general hybrid observable$A\in\gA$ is a function defined on the classical phase space
taking values on the set of quantum operators $\gaqm$.

The two natural generalisation of the Poisson brackets are due to
Anderson (1995),
\be
\{\![A,B]\!\}\defeq\{A,B\}_\hbar+\{A,B\}, \label{h95}
\ee
or  Alexandrov (1981) and Gerasimenko (1982),
\be
\{\![A,B]\!\}\defeq\{A,B\}_\hbar+\half\big(\{A,B\}-\{B,A\}\big). \label{ag81}
\ee
In both cases the quantum and classical Poisson brackets are evaluated on the relevant objects according to the standard rules. For $f,g\in\gacl$ and $A,B\in\gaqm$ the definition~\eqref{h95} result in
\be
\{\![fA,gB]\!\}=fg\{A,B\}_\hbar +AB\{f,g\},
\ee
and the expectation value of an operator $A(z)$ is given by
\be
\6A\9_\rho=\int\tr(\rho A)\mu(dz).
\ee

Applying the former version to the system of Eq.~\eqref{ex1}, we find the Hamiltonian equations of motion in agrement with both fully classical or fully quantum cases, perfectly satisfying the expectations dictated by the correspondence principle,
\begin{align}
\dot q=p, \qquad \dot p=-q-\lambda x, \label{eqq} \\
\dot x=k, \qquad \dot k=-x-\lambda q. \label{eqc}
\end{align}
However, for more general Hamiltonians the lack of antisymmetry (which indicates an obvious failure as a Lie bracket) leads to the possibility  that time-independent Hamiltonians do not commute, $\{\![H,H]\!\}\neq 0$, leading to energy non-conservation and non-positivity of $\rhcl$, as was already pointed out by (Anderson, 1995).

The linear antisymmetric  bracket of Eq.~\eqref{ag81} was designed to produce the standard-looking equation for the operator-valued density on the phase space, $\rho(x,k)$ via
\be
\pad_t \rho=-\{\![\rho,H]\!\},
\ee
which allows for conservation of the total energy $\6H\9=\mathrm{Tr}(H)=\int\!\tr(\rho H)\mu(dxdk)$.
Its Wigner function version is used for modelling   quantum rate
processes, such as proton and electron transport (Kapral, 2006; Crespo-Otero and Barbati, 2018). However it fails to be a Lie
bracket as the Jacobi identity is not fulfilled. Thus it lacks a Hamiltonian structure and leads to time-irreversible dynamics (Sergi et al., 2018).

 In fact, none of the combinations of the classical and quantum brackets that satisfy
\be
\{\![f,g]\!\}=\{f,g\}, \qquad \{\![A,B]\!\}=\{A,B\}_\hbar, \label{con:red}
\ee
for $f,g\in \gacl$ and $A, B\in \gaqm$ can fulfill all the common  properties of the classical and the quantum brackets. Indeed, under the standard rules (particularly, using the bracket's antisymmetry and the Leibnitz rule), one finds
\be
\{\![fA,gB]\!\}= \{f,g\}AB+fg[A,B]=\{f,g\}BA+fg[A,B],
\ee
which fails for non-commutative $A$ and $B$ (and  for coupling of any two quantization schemes with $\hbar_1\neq\hbar_2$, (Caro and Salcedo, 1999; Salcedo, 2012)).

Ideally, a hybrid scheme should satisfy all the requirements I-VII, and to this end the hybrid bracket properties should mimic those of its quantum and classical counterparts (Caro and Salcedo, 1999). Specifically, the hybrid bracket is a Lie  bracket, i.e. satisfies Eqs.~\eqref{ll} and \eqref{l2}. The antisymmetry of the bracket ensures that time-independent Hamiltonians are conserved.
The linearity guarantees that for two observables $F,Z\in\gA$ without an explicit time dependence their
linear combination is also free from the explicit time dependence. Compliance with the Jacobi identity guarantees this independence for the result of their bracket $\{\![F,Z]\!\}$. Satisfying the Leibnitz rule Eq.~\eqref{ll} ensures that
  the product of two  observables is consistent with time evolution the commutation relations among canonical variables, and the expression such as $[q,p]=i\hbar$ or $[x,k]=0$ are preserved.

  Eq.~\eqref{con:red} is necessary to satisfy III (independence of classical and quantum evolutions in absence of interaction), that should be supplemented by
  \be
 \{\![f,A]\!\}=0, \qquad f\in\gacl, A\in\gaqm.
  \ee
Hermiticity or reality of the relevant observers is preserved if
\be
 \{\![F,Z]\!\}^\dag=\{\![F^\dag,Z^\dag]\!\}
\ee
As we have seen that enforcing a general form of the Leibnitz rule is impossible, the minimal requirement is that a constant observables, such as $I$ do not evolve.  Demanding a weaker from of the Leibnitz rule,
\be
 \{\![f,gA]\!\}=\{f,g\}A, \qquad \{\![A,fB]\!\}=f\{A,B\}_\hbar
\ee
for $f,g\in\gacl$ and $A,B,\gaqm$, enforces this, as well as an independent evolution of each factor in a product observable if the two sectors
are dynamically uncoupled.

It was shown by Gil and Salcedo (2017) that if the Hilbert space of the quantum subsystem  is finite-dimensional, there is a unique hybrid bracket that satisfies the above requirements.  However, it
 does not preserve positivity of the resulting quantum density matrix (Requirement II).

A different perspective on the construction of a hybrid bracket was provided by Amin and Walton (2021). The bracket
\be
 \{\![F,Z]\!\}=\frac{1}{i\hbar}(F\ast Z-Z\ast F),
\ee
is defined with the help of an unspecified non-commutative associative product $\ast$ that acts on the algebra of operator-valued phase space functions $A(x,k)\in \gA$. Such bracket satisfies both the Jacobi identity and the Leibnitz rule. Taking the partial classical limit using the phase space representation of quantum mechanics (Sec.~\ref{ps}), it is possible to show that
\be
\ast=1+\frac{i\hbar}{2}(\tP+\Sigma), \label{hyb1g}
\ee
where $\Sigma$ is some symmetric binary operation on classical variables. If such $\Sigma$ can be found, then
\be
\{\![A,B]\!\}\defeq\{A,B\}_\hbar+\half\big(\{A,B\}-\{B,A\}+A\Sigma B-B\Sigma A\big). \label{hybgen}
\ee
It reduces to the bracket of Eq.~\eqref{ag81} for $\Sigma=0$.

Some specific constructions are discussed in Sec.~\ref{ps}. They fail as generators of the universal hybrid dynamics,  but indicate that the hybrid dynamics may be consistent for restricted types of the interaction terms, such as
\begin{align}
&\hin=A(q,p)(\alpha x+\beta k) \label{ic1}\\
&\hin=A(q,p)f(x), \qquad \hin=  A(q,p)g(k), \label{ic2}
\end{align}
where $A$, $f$, $g$ are arbitrary functions and $\alpha$, $\beta$ are real constants.

Conclusions of no-go theorems may be evaded if one of their premises is not realised in the proposed construction. This is the case of a scheme of Elze (2012),  that is constructed using the generalized Poisson bracket, where the role of canonical pairs is taken by $(\psi, i\hbar\psi^*)$ (Zhang and Wu, 2006). The resulting hybrid dynamics is equivalent to Eqs.~\eqref{zM} and \eqref{psM} (Salcedo, 2012), and thus shares the benefits and the drawbacks of the mean field models.

Another way to introduce the canonical variables and the Poisson bracket is given by Gay-Balmaz and Tronci (2022).

\section{Hilbert space models}\label{hspace}

A different way of avoiding the no-go results of Sec.~\ref{hb}  is based on using the Hilbert space realisation of the classical mechanics. This representation of classical mechanics was    developed by   Koopman  and  von Neumann (see Reed and Simon (1972), Peres (1995), Dammeier and Werner (2023) for the details), and was applied to the problem of hybrid dynamics by Sherry and Sudarshan (1978). Classical states, i. e.,  probability densities on a phase space $\eP$, are described as classical wave functions $\phi(x,k)$, elements of the Hilbert space  of square-integrable complex-valued functions on the phase space,  $\eH_\rcl= \mathrm{L}^2\big(\eP,\mu(dxdk)\big)$. The hybrid dynamics is defined on the tensor  product $\eH_\rqm\otimes\eH_\rcl$. Bondar et al. (2019) and Gay-Balmaz and Tronci (2022) provide a historical introduction and a comprehensive list of references.

\subsection{Hilbert space classical  mechanics}

We illustrate the construction in the simplest possible setting of a single particle in one dimension. Eq.~\eqref{Liouville} can be rewritten as
\be
 i\pad_t\rhcl=\{iH,\rhcl\}\eqdef  \hat{\cL}\rhcl,
\ee
 where $\hat{\cL}$ is the Liouville operator, or Liouvillian.
  The Liouville density  is never negative, so we define
a classical wave function via
\be
\rhcl\eqdef|\phi|^2,
\ee
(in general the classical wave function  can have a complex phase), which  satisfies the Schr\"{o}dinger--Koopman equation with the Liouvillian in place of the Hamiltonian operator.
Under reasonable
assumptions about the Hamiltonian, the Liouvillian is an essentially self-adjoint operator on $\mathrm{L}^2\big(\eP,\mu(dxdk)\big)$ and generates a unitary evolution,
\be
\6\phi|\phi'\9=\int\!\mu(dxdk)\phi^*(x,k,t)\phi'(x,k,t).
\ee

It is possible to further mimic quantum theory by introducing commuting
  multiplicative operators $\hat{z}=(\hat{x}, \hat{k})$,
\be
\hat{x}\phi=x\phi,\qquad \hat{k}\phi=k\phi.
\ee
Then the shift operator is $\hat{p}_x\defeq-i\pad_x$ and the boost operator is $\hat{p}_k\defeq-i\pad_k$. They can be combined into a vector $\hbpi\defeq-i\nabla_z$. The shift operators
  are not observable, but determine the dynamics via the Liouvillian,
\be
\hat{\cL}
=\pad_k H\hat{p}_x-\pad_x H\hat{p}_k=\big(J\nabla H\big)\cdot \hat{\Pi} =\vH\cdot \hbpi
\ee
It is possible to introduce the Heisenberg picture, and obtain the equations via a variational formulation.
As pure states are actually phase space distributions, classical wave functions do not represent pure states. Nevertheless, it is possible to describe phase space measurements via positive-operator valued measures, and introduce, at least formally, classical entanglement between the subsystems (Terno, 2006). Despite these formal similarities it is important to note that the Liouvillian is not only not the energy
\be
\6H\9=\int\!\mu(dz)H\rhcl,
\ee
but may be unbounded from below. In fact, this  happens   for harmonic oscillator (Peres and Terno, 2001).

The minimal coupling method of a $U(1)$ gauge theory allows introduction of a covariant formulation to the variational formulation of the Koopmanian dynamics. Under the   transformation
\be
i\pad_t\to i\pad_t-\Phi(z),\qquad i\nabla\to i\nabla+\bca(z),
\ee
the covariant Liouvillian is
\be
\hbar\hat{\eL}_H\defeq\hbar\hat{\cL}+\Phi -\vH\cdot\bca.
\ee
One choice of gauge potential is
\be
\Phi=H/\hbar, \qquad \bca\cdot dz=p\cdot dq/\hbar,
\ee
where the 1-form $\bca\cdot dz$ is set to be the symplectic potential, as the symplectic form $\omega^{(2)}=\hbar d\bca$. Then the modified the Schr\"{o}dinger--Koopman equation becomes
\be
i\frac{\pad\phi}{\pad t}=i\{H,\phi\}-\hbar^{-1}L\phi,
\ee
where the Lagrangian $L=p\cdot\pad_p H-H$, whereas writing the classical wave function in the polar form $\phi=\sqrt{\rhcl}e^{iS/\hbar}$ leads to a suggestive pair of equations
\be
\frac{\pad\rhcl}{\pad t}+\{\rhcl,H\}=0, \qquad \frac{\pad S}{\pad t}+\{S,H\}=L.
\ee

 A different (called Liouville or the harmonic oscillator) gauge has the vector potential part
 \be
 \bca\cdot dz=\half(k\cdot dx-x\cdot dk).
 \ee
For homogenous quadratic Hamiltonians in this gauge $\Phi-\vH\cdot\bca=0$ and  $\hat{\cL}=\hat{\eL}_H$

Auxiliary quantities (de Gosson,  2005),
\be
\hbz_\pm\defeq J(\pm\hbar\hbpi-\bca), \qquad \bj\defeq \phi^*\hbz_+\phi
\ee
allow one to write the Liouvillian as
\be
\hbar \hat{\eL}_H=H-\vH\cdot \hbz_+,
\ee
and the Hamiltonian functional (Bondar et al. 2019; Gay-Balmaz and Tronci, 2022; 2023), as
\be
h=\hbar\int\!\phi^*\hat{\eL}_H\phi \mu(dz)=\int\!H(|\phi|^2+\mathrm{div}\bj)\mu(dz).
\ee
Identifying $h\equiv\6H\9$ leads to
\begin{align}
\rhcl=|\phi|^2+\mathrm{div}\bj=|\phi|^2-\mathrm{div}\big(J\bca|\phi|^2\big)+\hbar\{\phi^*,\phi\}.
\end{align}
The normalisation is not affected by $j$, but
\be
\6z\9=\int\!\rhcl\mu(dz)=\int\!\phi^*\hbz_-\phi\mu(dz).
\ee
While the Liouville density is not positive definite, its
sign is preserved in time since the Liouville equation is a characteristic equation.

An algebraic approach to the Koopmanian mechanics is presented by Morgan (2023).

\subsection{Hybrid dynamics}
The hybrid Hilbert space is constructed as a direct product of the quantum Hilbert space $\eH_\mathrm{qm}$, say $\mathrm{L}^2(\mbr^3,d^3q)$ of a single spinless particle, and the classical Koopman -- von Neumann Hilbert space $\eH_\mathrm{cl}$. Mathematical details, including the question of measures on these spaces are discussed in de Gosson (2005) and Dammeier and Werner (2023).

If the fully quantum Hamiltonian is $\hat{H}(\hat{x},\hat{k},\hat{q},\hat{p},\hat{s})=\hH_\rcl+\hH_\rqm+\hH_\rin$, where $\hat{s}$ stand for the discrete degrees of freedom, a natural extension of the Koopmanian formalism is the hybrid Liouvillian (Bondar et al., 2019) is
\be
\hat{\eL}_{\hat{H}}=\hat{H}-\nabla\hat{H}\cdot\hbz_+
\ee
(to simplify the subsequent expressions we absorbed $\hbar$ into the definition of the hybrid Liouvilian). The C and Q operators commute and the Jacobi identity is satisfied by construction. The Schr\"{o}dinger equation for the mixed wave function $\Upsilon(z,x)$ is
\be
i\hbar\pad_t\Upsilon=\hel_\hH\Upsilon.
\ee

There is a variational principle that preserves the energy invariant
\be
h=\6\Upsilon|\hel_\hH|\Upsilon\9= \tr\int\!\Upsilon^\dag\hel_\hH\Upsilon\mu(dz).
\ee
Equating it with the total density $h=\tr\int\hH\hat{\rho}\mu(dz)$  identifies the hybrid density operator as
\be
\hat\rho(q,q',z)=\Upsilon(q,z)\Upsilon^*(q',z)+\mathrm{div}\big(\Upsilon(q,z)\hbz_+ \Upsilon^*(q',z)\big). \label{rhomix}
\ee
(compare with Sec.~\ref{stoc}). However, this density operator does not possess a closed Hamiltonian equation. Its evolution has to be expressed in rather convoluted form in terms of $\Upsilon$. It takes a simpler form if the so-called exact decomposition of the wave function (Abedi et al 2010),
\be
\Upsilon(z,q,t)=\psi_z(q,t)\phi(z,t), \qquad \int\!|\psi_z(q,t)|^2dq=1
\ee
can be obtained. This factorization, upon making classical phases unobservable by a gauge principle, leads to a nonlinear hybrid theory, as shown by Gay-Balmaz and Tronci (2022).

 Alternatively (Peres and Terno  2001),  we can consider any hybrid interaction terms that allow  fulfilment of as many of the desiderata of Sec.~\ref{axioms} as possible. Highlighting explicitly the C, Q and CQ parts, we write this Koopmanian operator as
\be
\hek=\hH_\rqm+\hbar\hel_{H_\rcl}+\hek_\rin.
 \ee

This hybrid approach was successfully applied to describe interactions in simple measurement models  and it was anticipated that the Requirements I and II impose constraints on the admissible interaction terms (Sherry and Sudarshan, 1978). For example, for the quadratic Hamiltonian of Eq.~\eqref{ex1} using the harmonic oscillator gauge we have
\be
\hel_{\hH}=\half\hat{q}^2+\half\hat{p}^2+\hbar(\hat{k}\hat{p}_x-\hat{x}\hat{p}_k)-\lambda\hat{q}\hat{p}_x-\half\lambda\hat{q}\hat{x}.
\ee

However, neither this form of  the interaction term nor any $\hek_\rin$ can reproduce the identical classical and quantum equations of motion (Terno 2006). Indeed, having both $[\hat{p},\hek_\rin]=-\lambda \hat{x}$,
$[\hat{k},\hek_\rin]=-\lambda \hat{q}$, as well as having all C operators to commute with all Q operators, is incompatible with  the Jacobi identity for $\hat{p}$, $\hat{q}$, and $\hek_\rin$.

The   gauge $\Phi=0,\bca=0$ allows   minimal modifications to the equations for the observables. In this case $\hek_\rin =-\lambda\hat{q}\hat{p}_x$, and three of the equations \eqref{eqq},\eqref{eqc} remain unchanged, while
\be
\dot {\hat{p}}=-\hat{q}-\lambda \hat{p}_y
\ee
now has to be supplemented with the equations for the unobservable  $\hat{p}_x$ and $\hat{p}_y$. Their dynamics   remains decoupled from other variables but now drives the evolution of observables, leading to violation of the energy conservation (Peres and Terno, 2001; Ahmadzadegan et al., 2016).

The construction of Eq.~\eqref{rhomix}, on the other hand,  conservs energy by construction. However, for non-trivial gauges the combined density operator ${\varrho}(z)$ is not positive-definite and its sign is not preserved in time. While the quantum reduced density operator is positive semidefinite,
\be
\varrho=\int\! {\varrho}(z)\mu(dz)=\int\Upsilon(z)\Upsilon^\dag(z)\mu(dz),
\ee
the classical marginal may become negative,
\be
\rhcl(z)=\tr {\varrho}(z)=\tr\left(\Upsilon(z)\Upsilon^\dag(z)+\mathrm{div}\big(\Upsilon(z)\hbz_-\Upsilon^\dag(z)\big)\right).
\ee
Gay-Balmaz and Tronci (2020) identified an infinite
family of hybrid Hamiltonians preserving the initial sign of $\rhcl$, thus fullfilling the minimal set of requirement for the hybrids. One important system  a  classical oscillator coupled to a quantum
two-level system with possibly time-dependent parameters (Gay-Balmaz and Tronci, 2023; Manfredi et al. 2023). A good agreement with the fully quantum treatment  is found for a series of
study cases involving harmonic oscillators with linear and quadratic time-varying coupling. In all these cases the classical
evolution (starting with the appropriately  selected configurations that we discuss in Sec.~\ref{ps}) coincides exactly with the oscillator dynamics resulting from the
fully quantum description.

A mathematically rigorous procedure that was introduced by Dammeirt and Werner (2023) allows consistent hybrid dynamics for quasi-free operations (i. e. the resulting states are Gaussian, characterized by a matrix of expectations and variances) that include evolution under quadratic Hamiltonians, and also general types of noise. Describing operations as quantum channels, it allows treatment of many important quantum-informational tasks with continuous variables, such as preparation, measurement, repeated observation, cloning, teleportation, and dense coding.

The hybrid Hilbert space, again the direct product $\eH_\rqm\otimes\eH_\rcl$, is built as a representation space for the algebra of  quantum and classical observables $q,p,z$. An important distinction with previous approaches is that the symplectic structure on C is disregarded, and the classical phase space is treated simply  as a real vector space without additional structure.

It is possible to evaluate  the computational advantage that the Hilbert space hybrid scheme can give.  The difficulty in simulating large quantum systems   may be overcome by using  quantum simulation. In principle, it can be  implemented using quantum computers or simpler and more robust quantum analog devices (Georgescu et al., 2014; Preskill, 2018).  Gonzalez-Conde  and  Sornborger (2023) studied quantum simulations with
  a semiclassical subsystem  described in Koopman -- von Neumann formalism. In  many-body
particle simulations  the improvement in   qubit resources over fully quantum simulations scales only logarithmically with the ratio of typical actions of the subsystems, $S_\rcl/S_\rqm$. However, in case of fields the improvement is linear in this ratio.

\subsection{Statistical ensembles in configuration space}

The approach of Hall and Reginatto (2005) can be also   traced  to the wave function-based representation of classical mechanics, with the Madelung hydrodynamical model for the Schr\"{o}dinger equation (Peres, 1995) as the pattern to follow (Salcedo, 2012). This QC hybrid  is described by two real functions configuration space functions, $\varrho(x,q)$ and $S(x,q)$. In the purely quantum case these are defined via
\be
\psi(q)=\sqrt{\varrho(q)}e^{iS(q)/\hbar},
\ee
and satisfy the pair of the Madelung equations, given below in  in Eqs.~\eqref{mad1} and \eqref{mad2} with only the $q$ and $\nabla_q$ dependent terms. The function $\varrho$, being the probability density, is non-negative and normalized. When only the C system is present, the two functions define a classical mixed state
\be
\rho(x,k)=\varrho(x)\delta\big(k-\nabla S(x)\big).
\ee
The two functions again satisfy a pair of equations. The first one -- the continuity equation --- is identical to the quantum case, and the second one, due to absence of the terms proportional to $\hbar^2$ is the Hamilton-Jacobi equation.

Taking observables in C or Q cases to be functionals of $\varrho$ and $S$ it is possible to introduce the variational Poisson bracket and the Hamiltonian functional that generates the dynamics.  The same construction is applied to the interacting QC systems. For example, for
\be
H=\frac{k^2}{2M}+\frac{p^2}{2m}+V(x,q)
\ee
The equations that govern the two functions are
\begin{align}
&\pad_t\varrho=-\frac{1}{M}\nabla_x\cdot(\varrho\nabla_x S)-  \frac{1}{m}\nabla_q\cdot(\varrho\nabla_q S) \label{mad1}\\
&\pad_t S= -\frac{1}{2M}\big(\nabla_x S\big)^2-\frac{1}{2m}\big(\nabla_q S\big)^2+\frac{\hbar^2}{2m}\frac{\nabla_q^2 \varrho^{1/2}}{\varrho^{1/2}}-V. \label{mad2}
\end{align}
The Poisson bracket is defined as
\be
\{\cA,\cB\}=\int dxdq\left(\frac{\delta\cA}{\delta\varrho}\frac{\delta\cB}{\delta S}-\frac{\delta\cA}{\delta S}\frac{\delta\cB}{\delta \varrho}\right). \label{seb}
\ee
The observables are represented by their expectation values. In particular, a phase space function $f(x,k)$ and an operator $\hat{A}$ result in
\be
\cF=\int\!dxdq\varrho f\big(x,\nabla_xS\big), \qquad \cA=\int\!dx\6\Upsilon(x)|\hat{A}|\Upsilon(x)\9,
\ee
where the QC wave function $\6 q|\Upsilon(x)\9=\sqrt{\varrho(x,q)}e^{iS(x,q)/\hbar}$ satisfies a nonlinear Schr\"{o}dinger equation
\be
i\hbar\frac{\pad\Upsilon}{\pad t}=\left(-\frac{\hbar^2}{2M}\nabla_x^2-\frac{\hbar^2}{2m}\nabla_q^2+V+\frac{\hbar^2}{2M}\frac{\nabla_x^2|\Upsilon|}{|\Upsilon|}\right)\Upsilon.
\ee

 Using the canonical pair $(\Upsilon, i\hbar\Upsilon^*)$ it is possible to show that the scheme has a Lie bracket that is defined on the set of observables, and it reduces to the Poisson bracket and the commutator for purely C and Q systems, respectively. The Ehrenfest relations generalize to hybrid systems, and in particular
the expectation values for the position and momentum observables of linearly coupled classical and quantum oscillators obey the classical equations of motion (Hall, 2008; Hall and Reginatto, 2005). However, beyond the usual issues that are brought by nonlinearity (potentially conflicting with some clauses of Requirement V),   the bracket \eqref{seb}
of a general  purely quantum observable with a general purely classical observable is not zero. As this result remains valid for $\hin\equiv 0$, it violates Requirement III.

\section{Phase space models}\label{ps}

The phase-space formulation of quantum mechanics provides
   an alternative way of analyzing hybrid quantum--classical
systems. In this formulation,   quantum and classical systems are described using  functions on the phase space. Quantum states are described by their Wigner functions, and the exact quantum dynamics is obtained if the
Poisson bracket is replaced with the Moyal bracket (Zachos et al., 2005). It is a convenient setting to study the classical limit (Peres, 1995; Landsman, 2017) and its partial version that is used to derive the hybrid schemes (Caro and Salcedo, 1999; Di\'{o}si et al., 2000; Amin and Walton, 2021).  We first describe this representation and then present the main features of the phase space hybrid dynamics.

\subsection{Phase space quantum mechanics}\label{ps1}

A (Weyl-ordered) operator $\hat{A}$ on $\eH=\mathtt{L}^2(\mbr^n)$ is represented as a phase space function $A(q,p)$ on $\eP=\mbr^{2n}$ via the Wigner transform (Hillary et al., 1984; Zachos et al. 2005; Schlosshauer, 2007),
\be
A(q,p)=\frac{1}{(2\pi\hbar)^n}\int\! dy e^{i p y/\hbar}\6q+\half y|\hat{A}|x-\half y\9.
\ee
Under this transformation the operator product on the Hilbert space is mapped into the $\star$-product of the phase space functions,
 \be
 \star\defeq \exp \left[\frac{i\hbar}{2}\left(\overset{\leftarrow}{\partial_x}\overset{\rightarrow}{\partial_k}-\overset{\leftarrow}{\partial_k}\overset{\rightarrow}{\partial_x}\right)\right]
 \equiv \exp\left( \frac{i\hbar\mathtt{P}}{2} \right), \label{moy}
 \ee
and the commutator maps to the Moyal bracket
\begin{align}
[\hat{A},\hat{B}]\to[\![A,B]\!]&\defeq \frac{1}{i\hbar}(A\star B-B\star A) \nonumber \\
&=\frac{2}{\hbar}A\sin\left(\frac{\hbar\mathtt{P}}{2}\right)B.
\end{align}
Expansion of Eq.~\eqref{moy} shows that  the Moyal bracket ia equal to the Poisson bracket plus  correction terms,
\be
[\![A,B]\!]=\{A,B\}+\mathcal{O}(\hbar^2). \label{mOh}
\ee
It is   easy to see that for quadratic functions,
the Moyal bracket coincides with the Poisson bracket.

The Wigner transform of a density operator results in the Wigner quasi-probability distribution, $W_\rho$ (Sec.~\ref{qm}).  Its dynamics is governed by the quantum counterpart of Eq.~\eqref{Liouville}
 \be
\pad_t W_\rho(q,p)=-[\![W_\rho,H]\!]. \label{LM}
 \ee

The question of equivalence of quantum and classical descriptions makes sense in the following context. A positive initial Wigner function $W(x,k,t=0)$ that corresponds to the quantum state $\hat\rho(t=0)$ can be identified with the Liouville function, $\rhcl(t=0)\leftarrow W(t=0)$. This function is evolved classically by Eq.~\eqref{classE}, and then the reverse identification is made: $W(t)\leftarrow \rhcl(t)$. If this represents a valid quantum  state $\rhqm(t)$ the procedure is consistent. If, furthermore, the phase space expectation values, calculated with $\rhcl(t)$ or, equivalently, the quantum expectations calculated with $W_{\rhcl(t)}$ are the same as the expectations that are obtained with the quantum-evolved state $\rhqm(t)$, the two descriptions are equivalent (Ahmadzadegan et al. 2016).

\subsection{Hybrid dynamics}
The first step in devising a phase space hybrid dynamics is to represent the entire system on the combined phase  space $\eP$ with the coordinates $(q,p,x,k)$.
The usual methods of reaching the classical limit (Zurek, 2003; Schlosshauer, 2007; Landsman, 2017), such as the use  of   coherent states (Di\'{o}si et al., 2000),  Moyal brackets (Caro and Salcedo, 1999) or their counterparts for various operator orderings (Amin and Walton, 2021) are adapted to taking the partial classical limit (over the variables $x,k$) and result in various hybrid schemes.  It is obtained by keeping only terms up to the  order $\hbar$ in the derivatives of $x$ and $k$ in the  $\star$-product.

The result for an arbitrary quantization scheme in the C subsystem (and thus a general  $\star$-product) give the explicit form of the phase space representation of the hybrid bracket  Eq.~\eqref{hybgen},
\begin{align}
\{\![A,B]\!\}&\defeq   [\![A,B]\!]_\mathrm{Q}+\half (A\!\star_\mathrm{Q}\!\tP B+B\!\star_\mathrm{Q}\!\tP A) \nonumber \\
&+\half (A\!\star_\mathrm{Q}\!\Sigma B-B\!\star_\mathrm{Q}\!\Sigma A),
\end{align}
where general first-order symmetric bidifferential operator $\Sigma$ (Amin and Walton, 2021) is given by
\be
\Sigma=a\overset{\leftarrow}{\partial_x}\overset{\rightarrow}{\partial_x}+b\overset{\leftarrow}{\partial_k}\overset{\rightarrow}{\partial_k}+c\big(\overset{\leftarrow}{\partial_x}\overset{\rightarrow}{\partial_k}+
\overset{\leftarrow}{\partial_k}\overset{\rightarrow}{\partial_x}\big),
\ee
where $a$, $b$, $c$ are real constants.
The $\star_\mathrm{Q}$-product and the associated bracket act nontrivially only on the Q variables $(q,p)$.

Different sets $(a,b,c)$   define different operators $\Sigma$ and different forms of hybrid dynamics (Amin and Walton, 2021).   For
example, the Weyl ordering  results in the set $(0,0,0)$ and thus in the bracket of Eq.~\eqref{ag81},   putting the no-go result of Caro and Salcedo (1999) in a wider context. The standard and antistandard orderings result in $(0, 0,\pm1)$, and the Husimi ordering with $(1,1,0)$ corresponds to the scheme of Di\'{o}si et al. (2000).

As   already discussed, none of the resulting schemes are universally consistent. Nevertheless, they are expected to work for restricted classes of interactions, such as those of Eqs.~\eqref{ic1} and \eqref{ic2}. We now consider some of the cases and point out the resulting restrictions and obstructions.

In this schemes the Wigner distribution $W(q,p,x,k,t)$ has as its marginals
\be
\rhcl(x,k,t)=\int\!Wdqdp, \qquad W_\rqm(q,p,t)=\int\!Wdxdk,
\ee
a valid classical Liouville density and a quantum Wigner function, respectively.

A hybrid system with the Hamiltonian \eqref{ex1} was studied on the phase space by Barcello et al. (2012). In this case the Moyal and the Poisson brackets are identical and there are no dynamical inconsistencies. However, the anticipated back reaction   not only introduces   quantum uncertainty to the classical system, but also reduces the quantum uncertainty. Taking initial pure classical states $\rhcl=\delta(x-x_0)\delta(k-k_0)$ (or, in general, classical probability distributions with $\Delta x\Delta k<\half\hbar$), leads to violations of the Heisenberg uncertainty relations for the quantum system.

A consistent hybrid scheme in this case is formalised by the epistemically-restricted Liouville mechanics that was proposed by Bartlett et al. (2012) to study the relations between quantum and classical onthology.  Their conclusion was that the underlying quantum mechanics, if restricted to the Gaussian operations on  Gaussian states can be mimicked by dynamics of  phase-space distributions, subject to two restrictions. These restrictions are   the classical uncertainty relation   and   the maximum entropy principle. The former implies that the covariance matrix of the probability distribution $\bm{\chi}$ must satisfy the inequality
\be\label{CUP}
\bm{\chi}+i \epsilon J/2\geq 0,
\ee
where to reproduce the Gaussian quantum--classical mechanics we have to set $\epsilon=\hbar$.  The latter condition requires that the phase-space distribution of the covariance matrix $\bm{\chi}$ has the maximum entropy compared to all the distributions with the same covariance matrix. Any distribution that satisfies these two conditions is a valid epistemic state and  can be equivalently described by a Gaussian state: its classical evolution on the phase space corresponds to the quantum evolution on the Hilbert space.
An immediate conclusion is that the hybrid quantum-classical description of the Gaussian states and operations is  possible at the price of accepting the unavoidable classical uncertainty.

Comparison of the quantum evolution equations for $W(q,p,x,k)$ that is based on the Moyal bracket Eq.~\eqref{LM} and the above (Moyal) hybrid bracket shows that already in the case of
\be
\hin=\beta q x^2,  \label{ex12}
\ee
 the hybrid and the full quantum evolutions are different. For this interaction the hybrid evolution equation for $W$ coincides with   the classical Liouville equation. Thus, this model satisfied the Requirements I, III--VII. However, even if this $\hin$ is a special case of Eq.~\eqref{ic2}, the hybrid dynamics is not consistent. It fails Requirement II, as can be seen from the study of
Ahmadzadegan at el. (2016).

To this end it is sufficient to follow the time evolution of various statistical moments (Ballentine and McRae, 1998; Brizuela, 2014), such as
\be
M^{a,b}_\mathrm{C}\equiv \lan \delta  {x}^a \delta  {k}^b \ran, \qquad M^{a,b}\equiv \lan \delta \hat{q}^a \delta \hat{p}^b \ran_{\text{ord}},
\ee
where the subscript `ord' refers to a particular ordering.
Here $\delta x=x-\lan x \ran$ and $\delta k=k-\lan k \ran$ are deviations from the mean values of position and momentum, respectively, in the classical system, and $\delta \hat{q}=\hat{q}-\lan \hat{q}\ran$ and $\delta \hat{p}=\hat{p}-\lan \hat{p}\ran$ are the operators for deviations from the mean (expectation) values.

For a particular  initial QC Gaussian state with non-zero QC correlations the evolution of $W(q,p,x,k)$ leads to violation of the Heisenberg uncertainty relation for $q$ and $p$ after some evolution time $t_*$. Hence
$W_\rqm(q,p,t_*)$ does not represent a valid quantum state $\hat{\rho}_\rqm(t)$, and the hybrid scheme breaks down.

\section{Stochastic dynamics}\label{stoc}

A view that all (with a possible exception of gravity), classical systems are just effective descriptions of the underlying quantum systems, as well as difficulties of the known   schemes that are based on the interaction of only Q and C subsystems, suggest adding to the discussion a mechanism that is responsible for the classicality of C (Di\'{o}si and Halliwell, 1998; Di\'{o}si et al. 2000).   Its key element is decoherence, which from the path integral perspective manifests itself as  washing out
interference between histories of certain types of variables. Mathematically it is effected
by some kind of coarse-graining procedure, such as coupling to a thermal bath or a measuring device, with subsequent tracing out of those degrees of freedom (Zurek, 2003; Schlosshauer, 2007; Busch et al., 2016).

Di\'{o}si and Halliwell (1998) considered a (quasi-)classical particle $M$ that is coupled to a quantum harmonic oscillator via $\hin=\lambda {q}x$. It is assumed that the classical particle continuously measures the
quantum one and is coupled to the momentarily measured value $\bar{q}$. With quantum state taken to be pure and normalized at each time step, the resulting stochastic system of equations consists of
\begin{align}
 i\hbar\pad_t\psi&=\big(\hH_\rqm+\lambda xq\big)\psi \nonumber \\
 &+\frac{i\hbar}{2\sigma}\left(\frac{1}{2\sigma}(q-\6 q\9)^2+(q-\6q\9)\eta(t)\right)\psi,
\end{align}
for Q and
\be
M\ddot{x}=-V'(x)-\lambda\6 q\9-\lambda\sigma\eta(t),
\ee
where $\6 q\9=\6\psi|\hat{q}|\psi\9$, parameter $\sigma$ represents the effective resolution of the measurement (and needs to scale as $\lambda^{-1}$, and $\eta(t)$ is the Gaussian white noise with zero mean and delta correlation function. This system gives intuitively sensible results even when the quantum oscillator starts as a superposition
superposition of well-separated localized states.

This scheme is non-linear, but this is not a problem for a purposely effective description. Two other drawbacks are the perpetual purity of the quantum state and the impossibility of describing separated Q and C systems as the coupling $\lambda\to0$.

A description of an idealized measurement (with $H_\rin=\lambda(t)x q$ assumed to be impulsive, so the free dynamics of the classical apparatus and the quantum system can be neglected), was developed by Milburn (2017).  The initial product state $\varrho=\rho_\mathrm{cl}\rhqm$ evolves into a valid $\varrho'(x,k)$ that is a result of decoherence in the position basis, as well as the Hamiltonian misplacement in the momentum and the noise added to it. The unconditional evolution of the quantum state $\rhqm\to\rhqm '$ by a trace-preserving completely positive  (TPCP) map whose Kraus operators depend on the classical position $x$, and the Liouville density $\rhcl '$ is obtained as an average of position displacements.

Once such transformations are invoked there is no reason why one cannot construct a general hybrid dynamics as a QC counterpart (Di\'{o}si, 2014; Oppenheim, 2023) of the TPCP evolution of open quantum systems. First we introduce the QC density operator.  We take a cue from the so-called zero discord quantum-classical states of quantum theory (Ollivier and Zurek, 2001; Brodutch and Terno, 2010) that have a form $\varrho=\sum_z p(z)|z\9_\mathrm{cl}\6 z|\otimes\rho_\mathrm{qm}(z)$, where $|z\9_{\mathrm{cl}}$ form an orthonormal basis on $\eH_\mathrm{cl}$, $\rho_\mathrm{qm}(z)$ are density matrices on Q and the positive weights $p(z)$ sum up to one. The hybrid density operator is defined as
\be
{\varrho}\eqdef\int\!\mu(dz)\varrho(z)\defeq \int\!\mu(dz)\rhcl(z)\rhqm(z), \qquad  \mathrm{Tr}\varrho=1,
\ee
where, as before, $\rhcl(z)$ is the probability density on the classical phase space $\eP$ and $\rhqm(z)$ is the operator-values function of $z\in\eP$.

The most general TPCP evolution of the hybrid QC state is generalisation of the GKSL equation (Oppenheim, 2023; Oppenheim et al. 2023). It has the form
\begin{align}
\frac{\pad\varrho(z,t)}{\pad t}=&-\{\varrho,H\}_\hbar+d_0^{\alpha\beta}L_\alpha\varrho L_\beta^\dag-\half d_0^{\alpha\beta}[L_\beta^\dag L_\alpha,\varrho]_+ \nonumber \\
&+\sum_{n=1}^{n=2}(-1)^n \left( \frac{\pad^n}{\pad z_{i_1}\ldots \pad z_{i_n}}\right)(d^{00}_{n,i_1\ldots i_n}\varrho) \nonumber \\
&+\frac{\pad}{\pad{z_i}}\big(d^{0\alpha}_{1,i}\varrho L^\dag_\alpha\big)+\frac{\pad}{\pad{z_i}}\big(d^{ \alpha0}_{1,i}L_\alpha\varrho \big),
\end{align}
where the first line consists of the three terms of the GKSL equations. Here $d_0$ is the matrix of Linbladian couplings, $d_1$ encodes the strength of the CQ interaction, and $d^{00}_2$ represent the unavoidable diffusion in the classical phase spaces. The matrices should satisfy $2d_2^{00}\succeq d_1 d_0^{-1}d_1^\dag$,   and $(I-d_0d_0^{-1})d_1=0$, where $d_0^{-1}$ is the generalized inverse of the positive semi-definite matrix $d_0$.

Taking the system of coupled oscillators of Eq.~\eqref{ex1} as an example (with $d_1\equiv \lambda$), the evolution becomes
\be
\pad_t\varrho=-\{\![\varrho,H]\!\}-\half\kappa\{q,\{\varrho,q\}_\hbar\}_\hbar+d_2\frac{\pad^2\varrho}{\pad k^2}+\gamma\frac{\pad k\varrho}{\pad k},
\ee
where the hybrid bracket of Eq.~\eqref{ag81} that is explicitly given by
\be
\{\![\varrho,H]\!\}=\{\varrho,H_\rcl\}+\{\varrho,H_\rqm\}_\hbar-\half d_1\big(q\pad_k\varrho+\pad_k\varrho\, q\big),
\ee
where the decoherence and diffusion (that is dampened by a friction term with coupling $\gamma$) ensure that the QC density matrix remains positive. The three parameters should satisfy $d_2\geqslant d_1^2/\kappa$.

In general, if the dynamics is memoryless, there are two classes of these dynamics, one with finite sized jumps in the classical phase space and one which is continuous (Oppenheim et al., 2023).

\section{Summary}
We examined various unitary hybrid shemes. Despite practical utility of some of them, each has its flaws. On the other hand, stochastic schemes that inject noise into the system are more cumbersome but maintain  consistency. It remains to be seen if indeed a consistent unitary scheme is impossible, whether the decoherence found in stochastic methods is simply the price of merging incompatible ideas or holds greater physical meaning.

\section*{Acknowledgments}
I am grateful to my friends and colleagues Aida Ahmadzadegan, Denys Bondar, Aharon Brodutch, Laios Di\'{o}si,  Flaminia Giacomini, Viqar Husain, Robert Mann, Gerard Milburn, Jonathan Oppenheim, Asher Peres, and Cesare Tronci  for our numerous discussions and collaborations. Suggestions and critical comments of Flaminia Giacomini, Javier Gonzalez-Conde, Lorenzo Salcedo,  and Cesare Tronci have greatly improved this article.

\section*{References}

\noindent
 Abedi, A.,  Maitra, N. T.,   Gross, E. K. U., 2010. 
{\href{https://doi.org/10.1103/PhysRevLett.105.123002}{Phys. Rev. Lett. \textbf{105}, 123002}.

 \noindent
Ahmadzadegan, A., Mann, R. B., Terno, D. R., 2016. 
{\href{https://doi.org/10.1103/PhysRevA.93.032122}{Phys. Rev. A \textbf{93}, 032122}.

\noindent   Aleksandrov, I. V., 1981. 
{\href{https://doi.org/10.1515/zna-1981-0819}{Z. Naturforsch. \textbf{36A}, 902}.

\noindent
Alonso, J. L., Clemente-Gallardo J., Cuch\'{\i}, J. C., Echenique, P., Falceto, F., 2012. 
{\href{https://doi.org/10.1063/1.4737861}{J. Chem. Phys. \textbf{137} 054106}.

\noindent
Amin, A., Walton, M. A., 2021. 
{\href{https://doi.org/10.1103/PhysRevA.104.032216}{Phys. Rev. A \textbf{104}, 032216}.

 \noindent
Anderson, A., 1995. 
{\href{https://doi.org/10.1103/PhysRevLett.74.621}{Phys. Rev. Lett. \textbf{74}, 621}.

\noindent
Arnold, V. I., 1989. \textit{Mathematical Methods of Classical Mechanics}. Springer, Berlin.

\noindent
Ballentine, L. E., McRae, S. M., 1998. 
{\href{https://doi.org/10.1103/PhysRevA.58.1799}{Phys. Rev. A \textbf{58}, 1799}.

\noindent   Barcelo, C.,   Carballo-Rubio, R.,   Garay, L. J.   Gomez-Escalante, R., 2012. 
{\href{https://doi.org/10.1103/PhysRevA.86.042120}{Phys. Rev. A \textbf{86}, 042120}.

\noindent  Bartlett, S. D.,    Rudolph, T.,   Spekkens, R. W., 2012. 
{\href{https://doi.org/10.1103/PhysRevA.86.012103}{Phys. Rev. A \textbf{86}, 012103}.

\noindent Bayer, M., 2006. \textit{Beyond Born--Oppenheimer: Conical Intersections and Electronic Nonadiabatic Coupling Terms}. Wiley--Interscience, Hobaken, NJ.

\noindent  Bondar, D. I,   Gay-Balmaz, F.,  Tronci, C., 2019. 
{\href{http://dx.doi.org/10.1098/rspa.2018.0879}{Proc. R. Soc. A \textbf{475}, 20180879}.

\noindent
Boucher, W., Traschen, J., 1988. 
{\href{https://doi.org/10.1103/PhysRevD.37.3522}{Phys. Rev. D \textbf{37}, 3522}.

\noindent  Branford, D.,  Dahlsten, O. C. O.,    Garner, A. J. P., 2018. 
{\href{https://link.springer.com/article/10.1007/s10701-018-0205-9}{Found. Phys. \textbf{48}, 982}.

\noindent
 Brizuela, D., 2014. 
{\href{https://doi.org/10.1103/PhysRevD.90.085027}{Phys. Rev. D \textbf{90}, 085027}.

\noindent
Brodutch, A.,   Terno, D. R., 2010. 
{\href{https://doi.org/10.1103/PhysRevA.81.062103}{Phys. Rev. A \textbf{81}, 062103}.

\noindent Busch, P., Lahti, P., Pellonp\"{a}\"{a}, J.-P.,Ylinen, K., 2016. \textit{Quantum Measurement}. Springer,  Cham, Switzerland.

\noindent Caro, J., Salcedo, L. L., 1999. 
{\href{https://doi.org/10.1103/PhysRevA.60.842}{Phys. Rev. A \textbf{60}, 842}.

\noindent Crespo-Otero, R., Barbati, M. 2018. 
{\href{https://doi.org/10.1021/acs.chemrev.7b00577}{Chem. Rev. \textbf{118}, 7026}.

\noindent Dammeier, L.,  Werner, R. F., 2023. 
{\href{https://doi.org/10.22331/q-2023-07-26-1068}{Quantum \textbf{7}, 1068}.

\noindent de Gosson MA., 2005. 
{\href{https://doi.org/10.1088/0305-4470/38/42/007}{J. Phys. A: Math. Gen. \textbf{38}, 9263}.

\noindent
Di\'{o}si, L.,  Gisin, N., Struntz, W. T., 2000. 
Phys. Rev. A \textbf{61}, 22108.
{\href{https://doi.org/10.1103/PhysRevA.61.022108}{Phys. Rev. A \textbf{61}, 22108}.

\noindent
Di\'{o}si, L., Halliwell, J. J., 1998. 
{\href{https://doi.org/10.1103/PhysRevLett.81.2846}{Phys. Rev. Lett.  \textbf{81}, 2846}.

 \noindent
Di\'{o}si, L., 2014. 
{\href{https://doi.org/10.1088/0031-8949/2014/T163/014004}{Phys. Scr. \textbf{2014}, 014004}.

\noindent
Filip, R, Mi\v{s}ta Jr., L., 2011. 
{\href{https://doi.org/10.1103/PhysRevLett.106.200401}{Phys. Rev. Lett.  \textbf{106}, 200401}.

\noindent Fulde, P. 2012. \textit{Correlated Electrons in Quantum Matter}. World Scientific, Singapore.

\noindent Galley, T. D., Giacomini, F.,   Selby, J. H., 2023. 
{\href{https://doi.org/10.22331/q-2023-10-16-1142}{Quantum \textbf{7}, 1142}.

\noindent  Gay-Balmaz, G.,  Tronci, C., 2020. 
{\href{https://doi.org/10.1088/1361-6544/aba233}{Nonlinearity, \textbf{33}, 5383}.

\noindent Gay-Balmaz, G.,  Tronci, C., 2022. 
{\href{https://doi.org/10.3934/jgm.2022019}{J. Geomet. Mech. \textbf{14}, 559}.

\noindent Gay-Balmaz, G.,  Tronci, C., 2023. 
{\href{https://doi.org/10.1088/1751-8121/acc145}{J. Phys. A: Math. Theor. \textbf{56}, 144002}.

\noindent
Gerasimenko, V. I., 1982. 
{\href{https://link.springer.com/article/10.1007/BF01027604}{Theor. Math. Phys. \textbf{50}, 49}.

\noindent
Georgescu, I. M.,    Ashhab, S.,   Nori,  F., 2014. 
{\href{https://doi.org/10.1103/RevModPhys.86.153}{Rev. Mod. Phys.  \textbf{86}, 153}.

\noindent Gil, V., Salcedo, L. L., 2017. 
{\href{https://doi.org/10.1103/PhysRevA.95.012137}{Phys. Rev. A \textbf{95}, 012137}.

\noindent  Gisin, N., 1990. 
{\href{https://www.sciencedirect.com/science/article/abs/pii/037596019090786N?via%3Dihub}{Phys. Lett. \textbf{143A}, 1}.

\noindent Gonzalez-Conde, J., Sornborger, A. T, 2023. 
{\href{https://arxiv.org/abs/2308.16147}{arXiv:2308.16147}.

\noindent
Hall, M. J., 2008. 
{\href{https://doi.org/10.1103/PhysRevA.78.042104}{Phys. Rev A \textbf{78}, 042104}.

\noindent
Hall, M. J., Reginatto, M., 2005.  
{\href{https://doi.org/10.1103/PhysRevA.72.062109}{Phys. Rev. A \textbf{72}, 062109}.

\noindent
  Hillery, M.,   O’Connell, R. F.,  Scully, M. O.,   Wigner, E. P., 1984. 
{\href{https://doi.org/10.1016/0370-1573(84)90160-1}{ Phys. Rep. \textbf{106}, 121}.

\noindent Husain, V., Javed, I., Singh, S., 2022. 
{\href{https://doi.org/10.1103/PhysRevLett.129.111302}{Phys. Rev. Lett. \textbf{129}, 111302}.

\noindent Husain, V., Javed, I., Seahra, S. S.,  X N., 2023. 
{\href{https://arxiv.org/abs/2306.01060}{arXiv:2306.01060}.

\noindent Ingarden, R. S., Kossakowski, A.,  Ohoya, M., 1997.  \textit{Information Dynamics and Open Systems: Classical and Quantum Approach}. Kluwer, Dordrecht, Netherlands.

\noindent  Janotta, P.,  Hinrichsen, H., 2014. 
{\href{https://doi.org/10.1088/1751-8113/47/32/323001}{J. Phys. A: Math. Theor. \textbf{47} 323001}.

\noindent Kapral R., 2006. 
{\href{https://doi.org/10.1146/annurev.physchem.57.032905.104702}{Annu. Rev. Phys. Chem. \textbf{57}, 129}.

\noindent Kiefer, K., 2012. \textit{Quantum Gravity}, third ed. Oxford University Press, Oxford.

\noindent Landsman, N. P., 1998. \textit{Mathematical Topics Between Classical and Quantum Mechanics}. Springer, New York.

\noindent Landsman, N. P., 2017. \textit{Foundations of Quantum Theory: From Classical Concepts to Operator Algebras}. Springer, Cham, Switzerland.

\noindent Manfredi, G.,  Rittaud, A.,  Tronci C., 2023. 
{\href{https://doi.org/10.1088/1751-8121/acc21e}{J. Phys. A: Math. Gen. \textbf{56}, 154002}.

\noindent Mielnik, B., 1974. 
{\href{https://doi.org/10.1007/BF01646346}{Commun. Mat. Phys. \textbf{37}, 221}.

\noindent Milburn, G. J, 2017. in Barnett, S. M. (Ed.), \textit{Journeys from quantum optics to quantum technology},
{\href{https://doi.org/10.1016/j.pquantelec.2017.07.002}{Prog. Quant. Elect. \textbf{54}, 19}.

\noindent Moiseyev, N., 2017. 
{\href{http://dx.doi.org/10.1063/1.4973559}{J. Chem. Phys. \textbf{146}, 024101}.

\noindent Morgan, P., 2023. 
{\href{https://doi.org/10.1016/j.aop.2020.168090}{Ann. Phys. \textbf{44}, 168090}.

\noindent   Ollivier, H.  Zurek, W. H., 2001. 
{\href{https://doi.org/10.1103/PhysRevLett.88.017901}{Phys. Rev. Lett. \textbf{88}, 017901}.

\noindent Oppenheim, J., 2023. 
{\href{https://10.1103/PhysRevX.13.041040}{Phys. Rev. X \textbf{13}, 041040}.

\noindent Oppenheim, J., 2023.
{\href{https://doi.org/10.1038/s41467-023-43348-2}{Nature Com. \textbf{14},7910}.

\noindent Oppenheim, J.,   Sparaciari, C.,  \v{S}oda, B. T.,    Weller-Davies, Z., 2023. 
{\href{https://arxiv.org/abs/2203.01332}{arXiv:2203.01332}.

\noindent Penrose, R., 1996. 
{\href{https://doi.org/10.1007/BF02105068}{Gen. Relat. Gravit. \textbf{28}, 581}.

\noindent
Peres, A., 1989. 
Phys. Rev. Lett. \textbf{63}, 1114.
{\href{https://doi.org/10.1103/PhysRevLett.63.1114}{Phys. Rev. Lett. \textbf{63}, 1114}.

\noindent
Peres, A., 1995. \textit{Quantum Theory: Concepts and Methods}. Kluwer, Dordrecht.

\noindent
Peres, A., Terno, D. R., 2001. 
{\href{https://doi.org/10.1103/PhysRevA.63.022101}{Phys. Rev. A \textbf{63}, 022101}.

\noindent
Preskill, J., 2018. 
{\href{https://doi.org/10.22331/q-2018-08-06-79}{Quantum \textbf{2}, 79}.

\noindent
 Reed, M.,   Simon, B.,  1972. \textit{Functional Analysis}. Academic, New York.

\noindent
 Reed, M.,   Simon, B.,  1975. \textit{Fourier Analysis, Self-Adjointness}. Academic, New York.

\noindent
Salcedo, L. L., 2012. 
{\href{https://doi.org/10.1103/PhysRevA.85.022127}{Phys. Rev. A \textbf{85}, 022127}.

\noindent
Schlosshauer, M., 2007. \textit{Decoherence and Quantum-to-Classical Transition}. Springer, Berlin.

\noindent Sergi A., Hanna G., Grimaudo R., Messina A., 2018. 
{\href{https://doi.org/10.3390/sym10100518}{Symmetry \textbf{10}, 518}.

\noindent
Sherry, T. N., Sudarshan, E. C. G., 1978. 
{\href{https://doi.org/10.1103/PhysRevD.18.4580}{Phys. Rev. D \textbf{18}, 4580}.

 \noindent
Terno, D. R., 2006. 
{\href{https://doi.org/10.1007/s10701-005-9007-y}{Found. Phys. \textbf{36}, 102}.

\noindent   Weedbrook, C.,    Pirandola, S.,   García-Patr\'{o}n, R.,    Cerf, N. J.,    Ralph, T. C.,   Shapiro, J. H.,   Lloyd, S. 2012, 
{\href{https://doi.org/10.1103/RevModPhys.84.621}{Rev. Mod. Phys. \textbf{84}, 621}.

\noindent
Zachos, C. K.,  Fairlie, D. B.,    Curtright, T. L. (Eds.), 2005. \textit{Quantum Mechanics in Phase Space}, World Scientific, Singapore.

\noindent Zhang, Q., Wu, B., 2006. 
{\href{https://doi.org/10.1103/PhysRevLett.97.190401}{Phys. Rev. Lett. \textbf{97}, 19040}.

\noindent Zurek, W. H., 2003. 
{\href{https://doi.org/10.1103/RevModPhys.75.715}{Rev. Mod. Phys. \textbf{75}, 715}.

\end{document}